# Composite materials with uncured epoxy matrix exposed in stratosphere during NASA stratospheric balloon flight

Alexey Kondyurin, Irina Kondyurina, Marcela Bilek

Applied and Plasma Physics, School of Physics (A28), University of Sydney, Sydney, NSW 2006, Australia

**Abstract**

A cassette of uncured composite materials with an epoxy resin matrix was exposed in the stratosphere (40 km altitude) over 3 days. Temperature variations of -76…+32.5$^0$C and pressure up to 2.1 Torr were recorded during flight. An analysis of the chemical structure of the composites showed, that the polymer matrix exposed in the stratosphere becomes crosslinked, while the ground control materials react by way of polycondensation reaction of epoxy groups. The space irradiations are considered to be responsible for crosslinking of the uncured polymers exposed in the stratosphere. The composites were cured on Earth after landing. Analysis of the cured composites showed, that the polymer matrix remains active under stratospheric conditions. The results can be used for predicting curing processes of polymer composite in a free space environment during an orbital space flight.

**Introduction**

Future space exploration will require large light-weight structures for habitats, greenhouses, space bases, space factories and so on. A new approach enabling large-size constructions in space relies on the use of the technology of polymerization of fiber-filled composites with a curable polymer matrix applied in the free space environment. For example, a fabric impregnated with a long-life matrix (prepreg) can be prepared in terrestrial conditions and, after folding, can be shipped in a container to orbit and unfolded there by inflating. Then the matrix polymerization reaction is initiated producing a durable composite wall or frame. Using such an approach, there are no restrictions on the frame size and form of the construction in space and the number of deployment missions is kept at a minimum.
In free space, the material is exposed to high vacuum, dramatic temperature changes, plasma of free space due to cosmic rays, sun irradiation and atomic oxygen (in low Earth orbit), micrometeorite fluence, electric charging and microgravitation. The development of appropriate polymer matrix composites requires an understanding of the chemical processes of polymer matrix curing under the specific free space conditions to be encountered.
Our preliminary studies of the polymerization process in a high vacuum, space plasma and subject to temperature variations indicate that for specific prepeg preparations the polymerization process is likely to be successful in free space and that the composite cured in a free space environment will have satisfactory mechanical properties [1-17]. However, the curing processes are sensitive to free space factors such as high vacuum, flux of high energy particles and temperature variations encountered.
Particularly pertinent observations from our previous work include:

- The evaporation of the active components can stop the curing reaction and evaporation can cause bubble formation in the curing polymer matrix and compromise the mechanical properties of the cured matrix.
- High energy space irradiations can destroy macromolecules and create free radicals, which can accelerate the curing kinetics and strengthen the composite.
- Temperature variations change dramatically the curing kinetics and evaporation process.

Our studies show, that the curing process can proceed and a durable composite material can be polymerized under simulated free space conditions. However in a laboratory environment it is not possible to simulate accurately the combinations of factors observed in space in order to assess how the various influences couple. To develop the appropriate polymer matrix composition for use in a particular free space environment, the effects of the prevailing free space conditions acting together must be taken into account. More detailed investigations of the curing process under real free space conditions, where all these free space factors act simultaneously during the curing process are required.

The goal of the flight experiment was an investigation of the effect of the stratospheric conditions on the uncured polymer matrix of the composite material. Stratospheric conditions are expected to have a strong impact on chemical processes in polymer materials. The unique combination of low atmospheric pressure, high intensity UV radiation including short wavelength UV, diurnal temperature variations and other aspects associated with solar irradiation strongly influences the chemical processes in polymeric materials. Since such conditions cannot be adequately simulated in the laboratory, it is difficult to predict the impact on the curing chemistry, particularly important in designing polymers which could be shaped and cured in space for large scale structural applications.

The flight experiment with uncured composites was a part of the NASA scientific balloon flight program realised at the NASA stratospheric balloon station in Alice Springs, Australia. A flight cassette with the samples was installed on a 1200 kg payload carrying the telescope of the Tracking and Imaging Gamma-Ray Experiment (TIGRE). The payload was lifted with a "zero-pressure" stratospheric balloon filled with Helium. The dimension of the completely unfolded balloon is about 300 m. Columbia Scientific Balloon Facility (CSBF) provided the launch, flight telemetry and landing of the balloon and payload.

**Stratospheric flight**

The uncured composite prepreg with epoxy matrix SE70 (Gurit, Australia) filled with carbon fibers was purchased from Australian branch of Gurit. The prepreg was one layer of 100x50 mm2 size sheets and was kept in fridge (4C) before using. The epoxy resins Polypox E 375 (based on Bisphenol F and epichlorhydrine, UPPC AG), Aldrich (CAS 25036-25-3, Poly(Bisphenol A-co-epichlorohydrin), glycidyl end-capped, $M_n$=374), Epilox AF 8-50 (based on Bisphenol F and A, Leuna-Harze GmbH) and ED-20 (based on Bisphenol A, $M_n$=320-340, Dzerzhinsk) were used. The hardener triethanolamine TEA (CAS 112-24-3, Aldrich) was added to the resins in 1:10 weight ratio and mixed for 1 minute. The mixtures were prepared on the 29

March, 2010 in a chemical laboratory of St. Phillip College, Alice Springs. The prepared mixtures were kept in a fridge ($2^0$C) for 5 days before using. Then the mixtures were vacuumed 2 minutes up to foaming and placed on glass fabric of size 300x25 mm$^2$. The 30 weight percent content of polymer matrix on fabric was achieved after removing loose resin with filter paper sheets.

Three cassettes with uncured samples were prepared. The flight cassette consists of an aluminium base and a sleeve of black Polyether fabric. The aluminium base was covered with paint filled with ZnO particulars (Dulux, Australia). The first datalogger with temperature sensor and microprocessor data storage unit (EL-USB-1, model 23039-50, USA) was placed in an aluminium cylinder, sealed with aluminium disks and glued hermetically with Araldite epoxy resin (Fig.1). The cylinder was covered by uncured SE70 Gurit prepreg. The top of base was covered with two layers of 0.05 mm thick Low Density Polyethylene (LDPE) films.

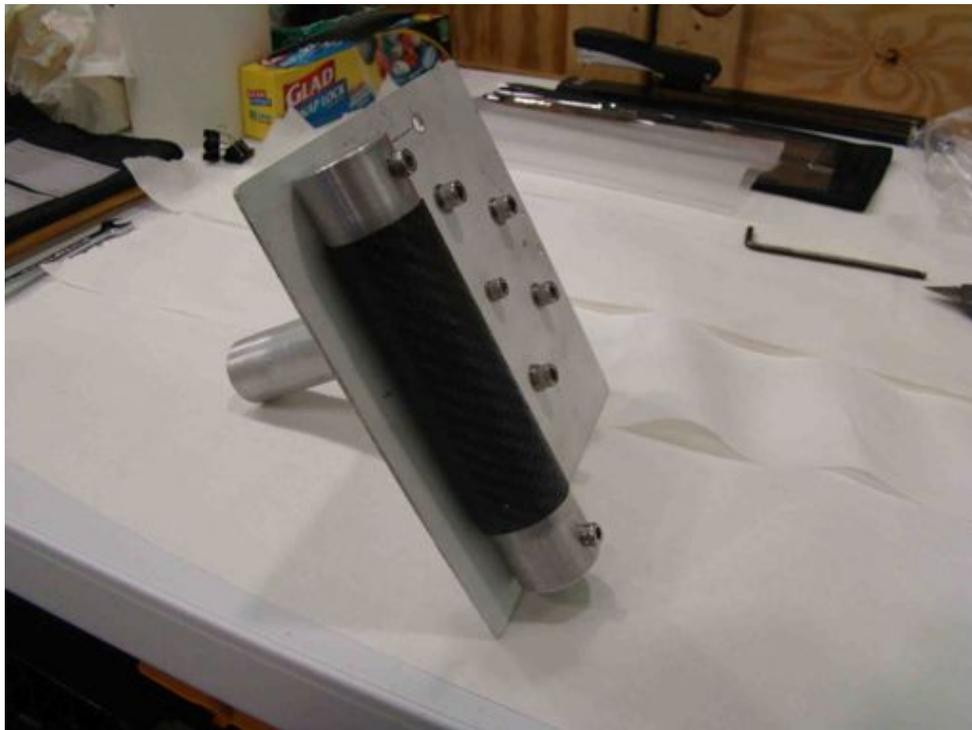

Fig.1. The base of the flight cassette: the cylinder with thermometer 1 inside and Gurit prepreg on it. A LDPE film is placed behind the cylinder and fixed with 7 screws. A second cylinder with thermometer 2 is under the base.

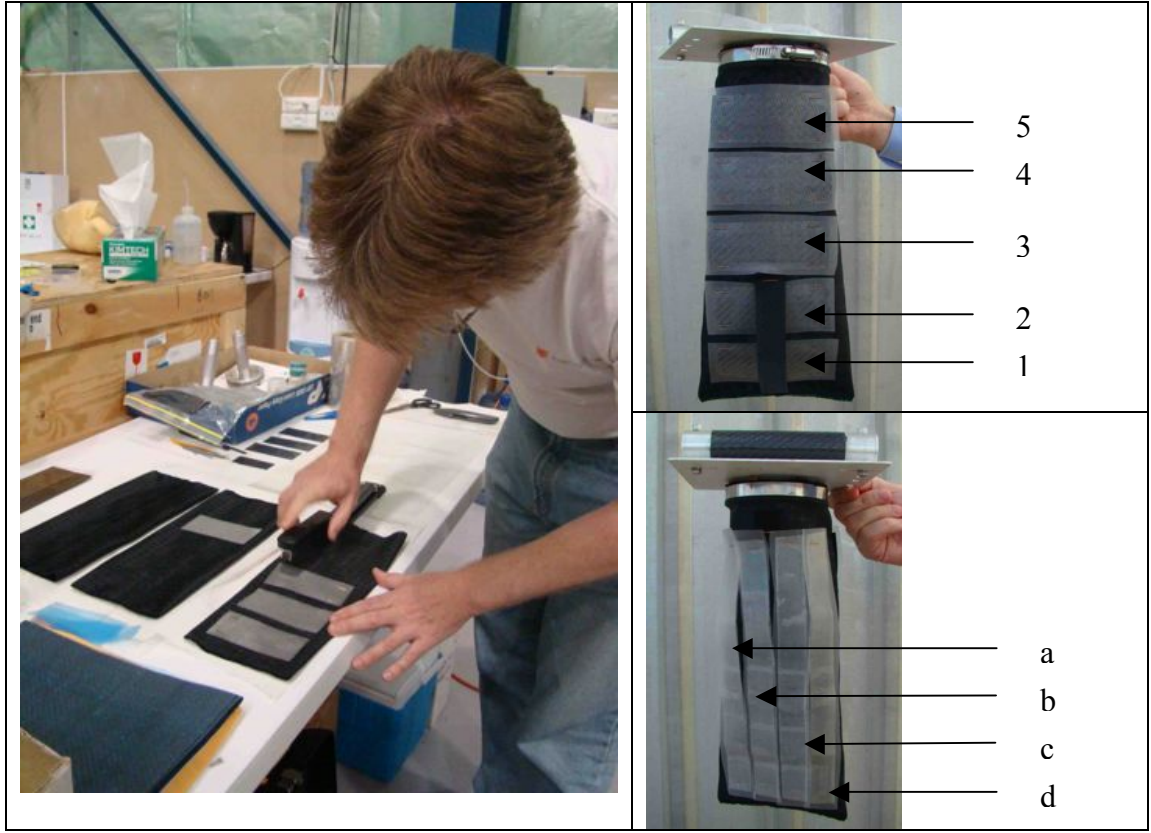

Fig.2. Preparation of the sleeve: stapling of the SE70 samples covered with Teflon mesh to the sleeve and the control cassette with the samples. (1-5) are SE70 prepregs, (a-d) prepregs with TEA hardener.

The 4 sheets of uncured prepregs of SE70 were covered by Teflon mesh of 0.04 mm cells and stapled to the sleeve (Fig.2). A sheet of SE70 prepreg was cured in thermobox at $80^0$C over 24 hours and stapled to the sleeve as a control sample (number 5 on Fig.2)). 4 strips of uncured Polypox (a), Aldrich (b), Epilox (c) and ED-20 (d) prepregs were covered by Teflon mesh and stapled along to the sleeve (Fig.2). 4 high-density polyethylene (HDPE) envelopes of size 200x50 mm$^2$ were filled to 10% of their expanded volume with residual air and hermetically sealed. These were then inserted into the open HDPE envelope and inserted into the sleeve. The aluminium cylinder with a second temperature datalogger was placed into a sleeve which was fixed to the base. The sleeve was folded 3 times and fixed by Velcro adhesive stapled to the sleeve. The total weight of the flight cassette was 1 kg.

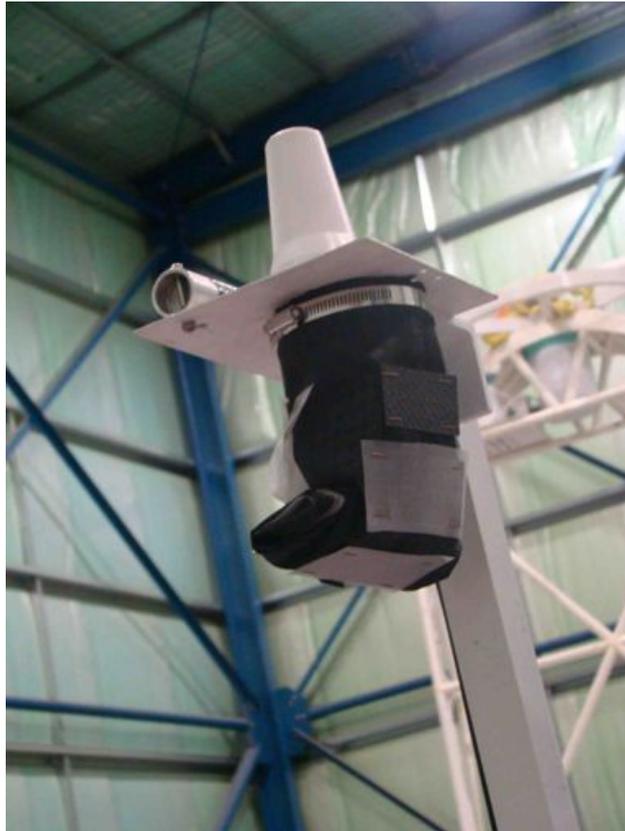

Fig.3. The flight cassette is fixed on the metal bar of the GPS antenna of the payload. The control sample, which appears dark in the picture, has been cured prior to the flight and is not covered with a Teflon mesh.

A ground control cassette was prepared at the same time with the same samples, excluding the cured SE70 prepreg, which was replaced by uncured prepreg. A third datalogger was placed in the cylinder and fixed on the base.

A fridge control cassette was prepared at the same time with the same samples without base and datalogger. The cassettes were prepared at 10 pm on 3 April, 2010. This time and all further references to time in the text refer to the local, Northern Territory time.

The flight cassette was fixed on the payload to the GPS antenna bar (Fig.3 and 4). The payload was moved out of the NASA base shed, moved to the airstrip and kept there from 1 am to 7 am on the 4 April, 2010. The launch was cancelled due to strong wind. The payload with cassette was moved back into the shed, the flight cassette was removed from the payload and both flight and ground control cassette were placed in a fridge (+2-3$^0$C).

The next day at 1 am the flight and ground control cassettes were taken out of fridge. The flight cassette was fixed to the payload again and moved out of the shed. After 5 hours, the cassette was removed from the payload and both cassettes were placed back into the fridge. The launch was cancelled due to strong wind (20 knots (10 m/sec)) at 400 m altitude since the maximum permitted was 10-12 knots (5-6 m/sec).

For the next 10 days the weather did not allow for the balloon launch: wind up to 50 knots (25 m/sec), thunderstorms and flooding. After the rain the soil had become too soft for the crane and heavy trucks with Helium. The drying and hardening of the soil took a further 2 days. The cassettes were kept in fridge during this period.

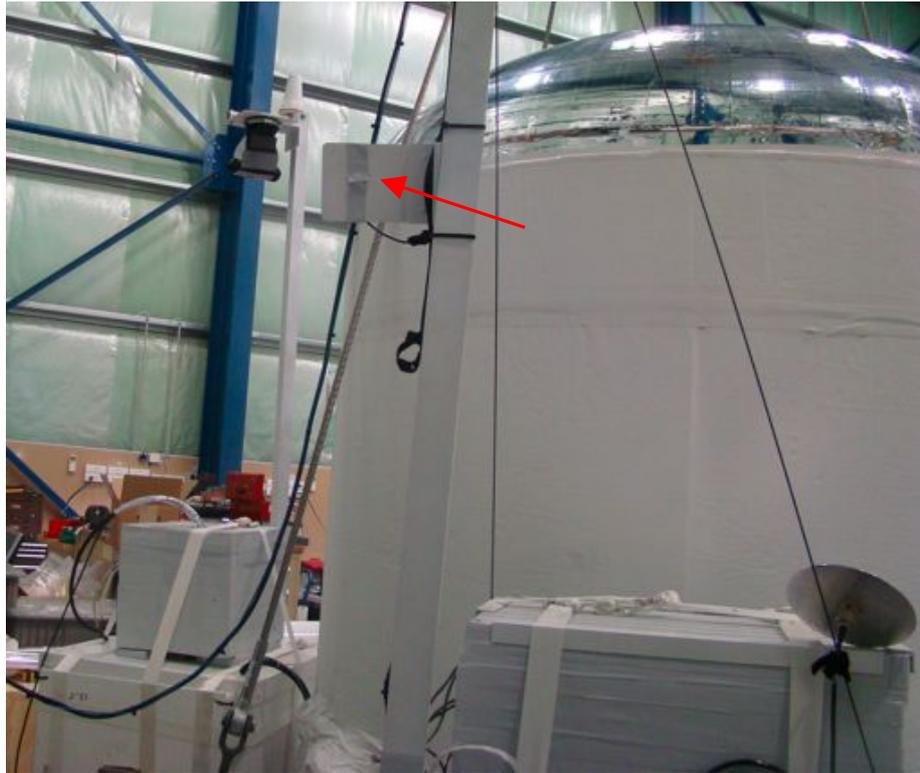
Fig.4. The array shows the back of the video camera. The camera was used to view the cassette during the flight.

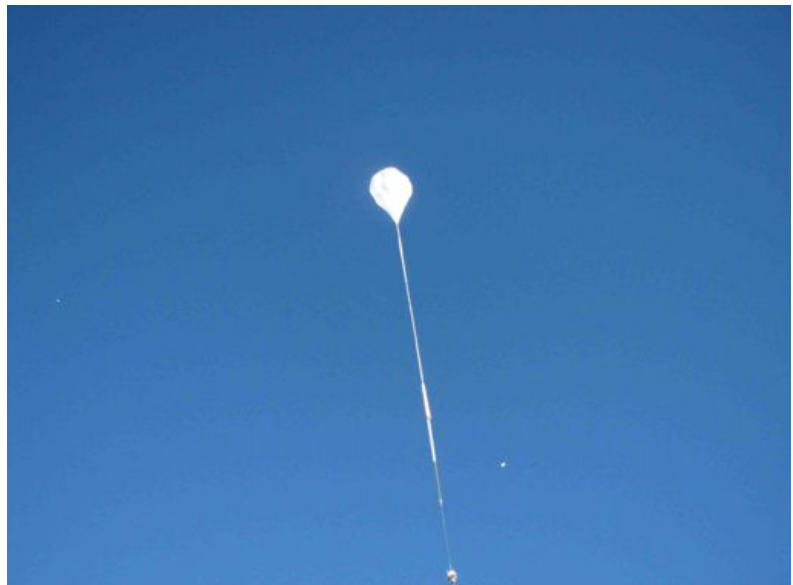
Fig.5. Balloon launch with payload.

At 1 am on the 16 April, the flight and ground control cassettes were taken from the fridge. The flight cassette was fixed on the payload and moved out to the airstrip. The balloon was launched from Alice Springs Seven Mile airport, Northern Territory, Australia, in the 16 April, 2010, at 9:00 (Fig.5). After 4 min at the altitude of 1.6 km the HDPE bags inside the sleeve expanded due to low pressure outside the sleeve, the Velcro adhesive peeled off and the flight cassette unfolded. After 2 hours the balloon rose to 40 km altitude.

Over the next three days the altitude of balloon varied between 40 km (day time) and 35 km (night time). The position of the balloon, altitude, pressure, temperature and signal from the video camera were monitored with telemetry (Fig.6). The payload was rotated by a motor at a rate of 1 turn per 4 minutes during two days of flight. On the 3$^{rd}$ day of flight, the rotation was stopped and the orientation of the payload was no longer controlled. After 3 days of flight, the payload was separated from the balloon and descended by means of a parachute. After 3 hours of descent the payload landed about 100 km to the west from Longreach, Queensland, Australia (latitude 24$^0$ 2.71' S, longitude 143$^0$ 54.5' E), 990 km from the launch site, at 18:11 on the 18 April, 2010. The speed at landing was about 4-5 m/sec. At landing, the payload fell on the side where the cassette was. The top of the cylinder was in contact with the soil. The day after landing the payload was found and transported to Longreach airport. The flight cassette was removed from the payload and stored in a fridge (+2$^0$C).

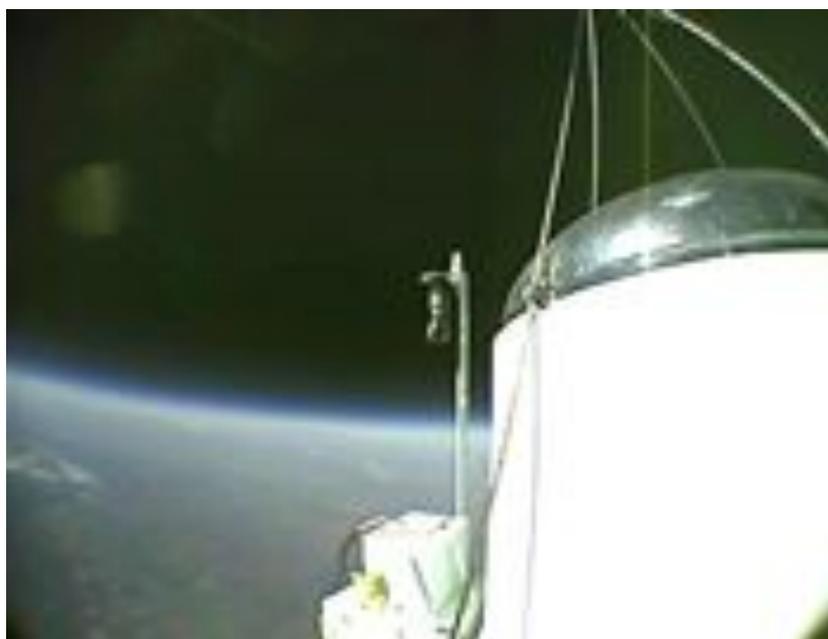
Fig.6. The view of the cassette sent by telemetry during the flight at 40 km altitude.

The ground control cassette was placed outside the NASA base shed at the time of launch and exposed to sun light during the balloon flight. As soon as the balloon landed, the ground control cassette was placed in enclosed in a dark container at stored at room temperature. The ground control cassette was placed in the fridge at the same time as the flight cassette was removed from the payload and placed in the fridge. The two cassettes were transported to Sydney in a fridge and stored in a fridge prior to analysis. Data loggers recorded the temperature on both cassettes during the waiting period before launch, the flight and during transportation.

**Measurements during the flight**

Pressure, temperature and altitude were recorded and logged during the flight. These data were sent to the ground station using telemetry. The air temperature was measured by a sensor placed in a white painted box at the bottom of the payload near the battery box. It showed, that the temperature decreased after launch from +20$^0$C on the ground reaching -76.7$^0$C at an altitude of 17600 m. During the flight at 40 km altitude, the temperature remained in the range -20 to +5$^0$C during the day and -30 to -45$^0$C at night (Fig.7 and 8).

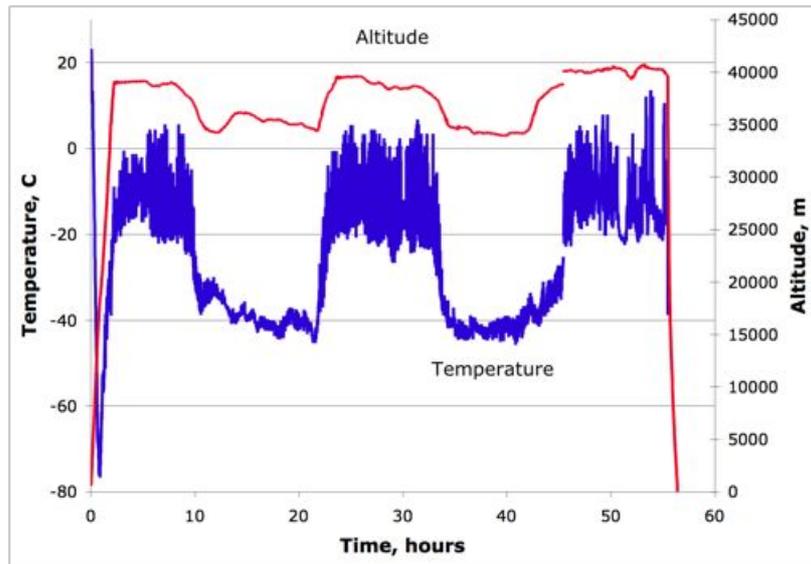

Fig.7. Temperature and altitude of the payload during the flight. Telemetry data. The break of the curve at 45 hours 40 min is due to change of the telemetry signal from the Alice Springs station to the Longreach station.

The daytime temperature varied periodically with time. The cycle of temperature variation corresponded to the rotation of the payload. Therefore, the high temperature corresponds to an orientation of the temperature sensor to the sun, when sunlight heated the sensor. The low temperature corresponds to the temperature of air.

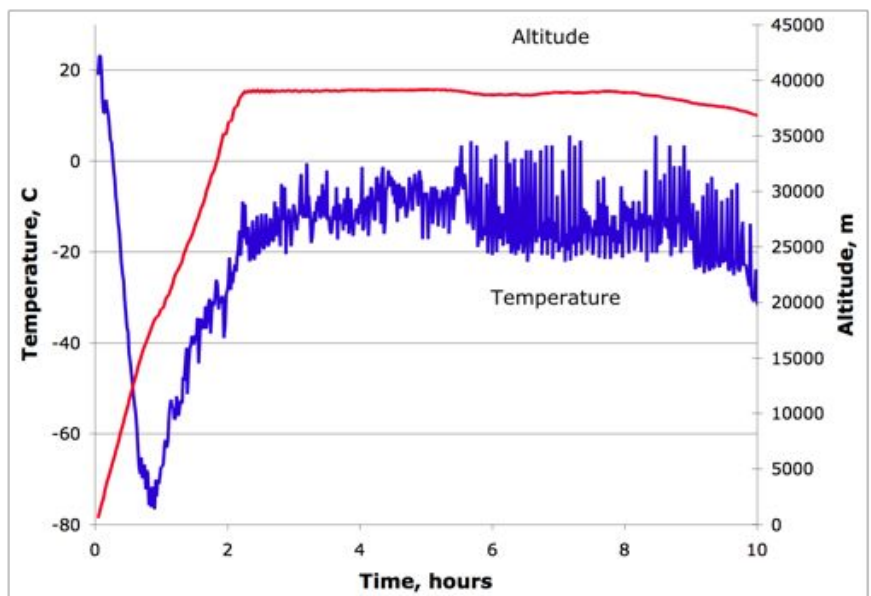

Fig.8. Temperature and altitude of the payload during the flight. Launch and first day data.

On the 3$^{rd}$ day, the rotation of the payload stopped and the temperature varied between -22 …+13$^0$C randomly depending on the orientation of the payload with respect to the sun (Fig.9).

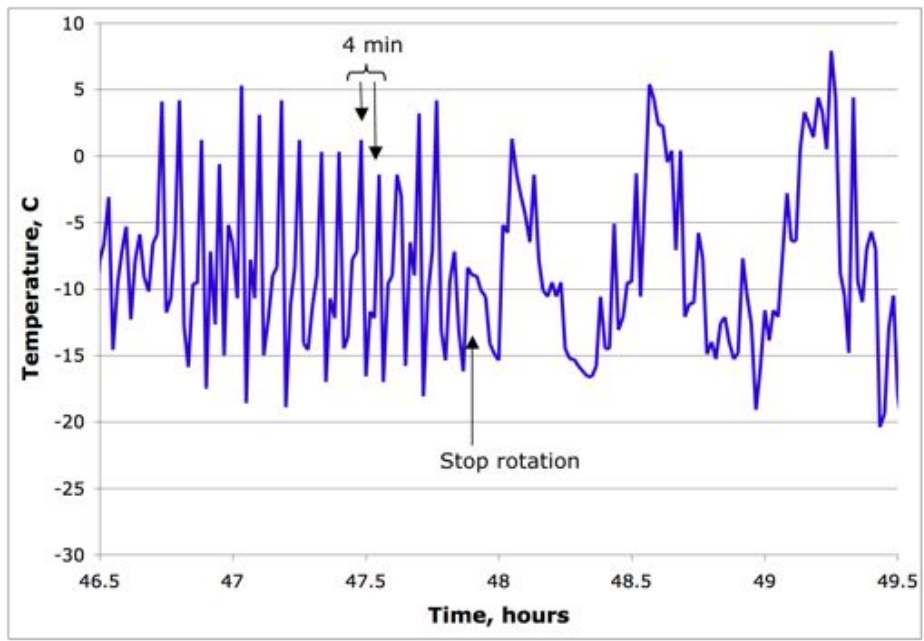

Fig.9. Variation of the temperature of the payload during the 3$^{rd}$ day of flight. The cycling of temperature up to 47 h 54 min is due to the rotation of the payload. After this time the payload was not rotated and the temperature variations correspond to its random orientation with respect to the sun.

The pressure decreased with elevation of the balloon (Fig.10). At daytime, the balloon was heated by solar irradiation. The altitude of the balloon remained in the range 38-39 km during the 1st and 2$^{nd}$ days of flight and then increased to 40 km on the 3$^{rd}$ day of flight. The pressure at daytime decreased to 2.5-2.1 Torr while at night time, the balloon was cooled, reducing the lift on the balloon and the altitude decreased to the range 34-35 km. Consequently, the pressure increased to 5 Torr at night time (fig.10 and 11).

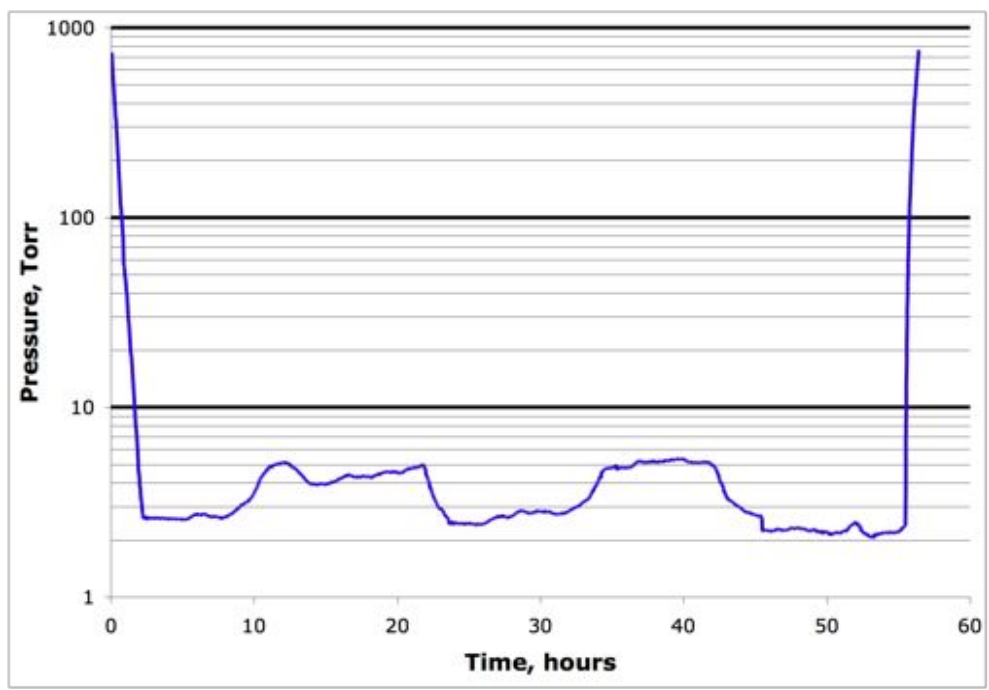

Fig.10. Air pressure near the payload as a function of time during the stratospheric flight.

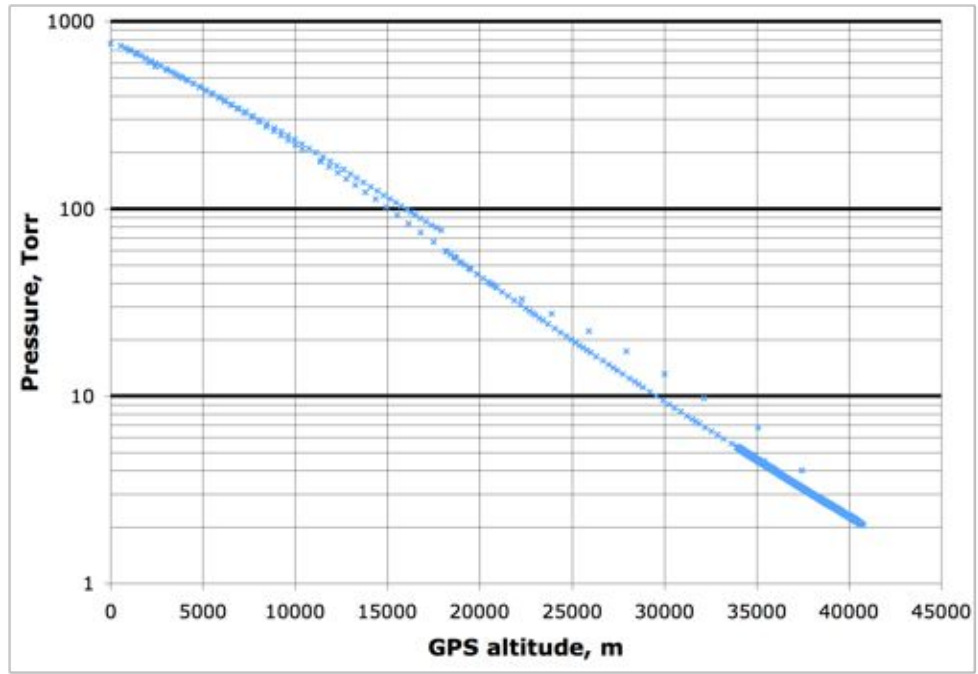

Fig.11. Variation of air pressure near the payload with altitude of the balloon.

Figure 12 shows the temperature measured inside the cylinder and sleeve of the flight cassette. At point "a", the cassette was taken from the fridge and attached to the payload. The cassette heated up to 22-23$^0$C which was the ambient temperature in the NASA base shed. Then the payload was moved out of shed and the temperature decreased to the air ambient temperature (8-9$^0$C). At point "b" the balloon was launched, the temperature decrease to -39$^0$C as the balloon rose. The temperature then increased to 26$^0$C in the cylinder and to 20.5$^0$C in the sleeve due to solar irradiation at higher altitude. At night time, the temperature decreased below -40$^0$C. Temperatures lower than -40$^0$C were not recorded because this is the lower limit of the measurement range for our thermometers.

At 7:25 am, on the second day, thermometer 2 in the sleeve stopped recording (point c). Perhaps, the battery froze. The maximum temperature of the flight cassette (+32.5$^0$C) was observed for on the 3$^{rd}$ day of the flight. The low (-38$^0$C) temperature observed on the 3rd day corresponds to a descent of the payload to an altitude of 20 km. At point "d", the payload landed.

The temperature of ground control cassette (3) was consistently higher than the flight cassette. The maximum temperature of the ground cassette was +37.5$^0$C.

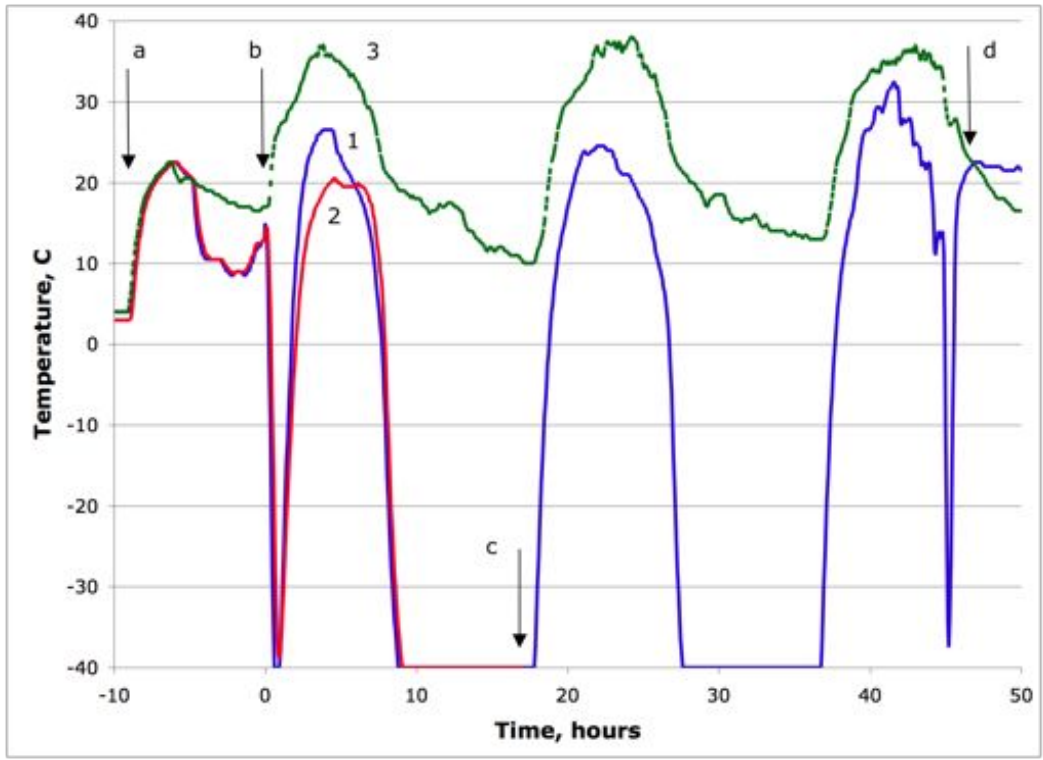

Fig.12. Temperature measured in the flight cassette (1, 2) and in the ground control cassette (3). Thermometer 1 was placed on the top of the base, thermometer 2 was placed in the sleeve of the flight cassette. Times marked are a –cassettes taken from the fridge, time b – launch, time c – thermometer 2 ceased recording, time d – landing of the payload.

The fridge control cassette was kept in fridge (2-3$^0$C) during the entire experiment prior to analysis.

**Analysis of the samples**

The samples from the flight cassette, the ground control cassette and the fridge control cassette were analysed. Progress of the curing reaction was monitored with Fourier Transform Infrared (FTIR) Attenuated Total Reflectance (ATR) spectroscopy. The degree of crosslinking was assessed by gel-fraction and swelling measurements.

The swelling degree of the samples in acetone was calculated according to

$$S=(M_1-M_f)/(M_2-M_f) \qquad (2)$$

where $M_1$ is weight of the sample swelled in acetone after soaking for 3 days in acetone but prior to drying. $M_2$ is the weight of the sample after soaking for 3 days in acetone and then drying. $M_f$ is the weight of the carbon fibres, calculated according to the known content of fibres in the original sample.
The gel-fraction was measured by dissolving of the uncured epoxy matrix in acetone. The gel-fraction content was calculated as

$$G=(M_2-M_f)/(M_3-M_f) \qquad (1)$$

where $M_3$ is weight of the sample before exposure to solvent.

The gel-fraction and swelling degree for each sample sheet was measured for 4 pieces (15x20 mm$^2$) cut from the sheet and the results were averaged. The first measurement of the gel-fraction was performed on the 24 of April, 6 days after landing. The second measurement of the gel-fraction was performed on the 5$^{th}$ of May, 17 days after landing. The samples were kept in fridge at +2-3$^0$C prior to all measurements.

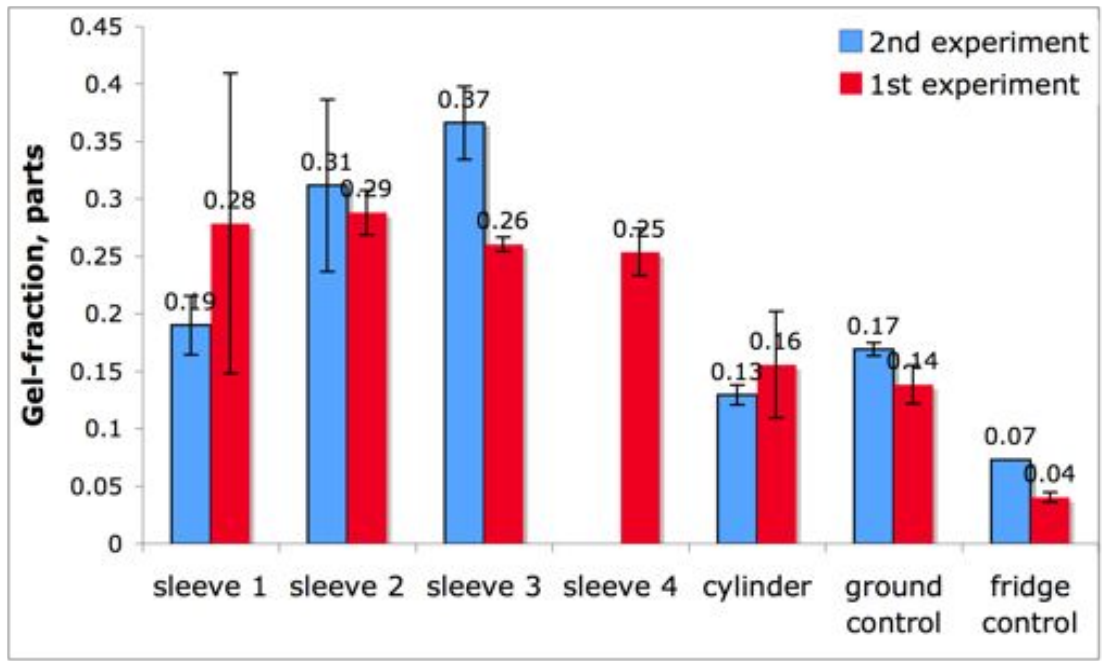

Fig.13. Gel-fractions of uncured SE70 prepreg samples in acetone. 1-4 sleeve sheets and the sheet from the cylinder are flight samples. The control samples are from the ground cassette and from the cassette stored in the fridge. The 1$^{st}$ measurement of the gel-fraction measurement was done 6 days after landing. The 2$^{nd}$ measurement of the gel-fraction was done 17 days after landing. The samples after landing were kept in the fridge at 2-3$^0$C prior to all measurements.

The gel-fraction content in the flight samples on the sleeve is higher, than in the ground control samples and the fridge control samples (Fig.13). The gel-fraction content in the flight sample on the cylinder is the same as in the ground control sample. The error bars of gel-fraction content in the flight sleeve samples are significantly higher than in the ground and fridge samples. These differences in flight and control samples are observed in the first measurement after 6 days after landing and in the second measurement after 17 days after landing.

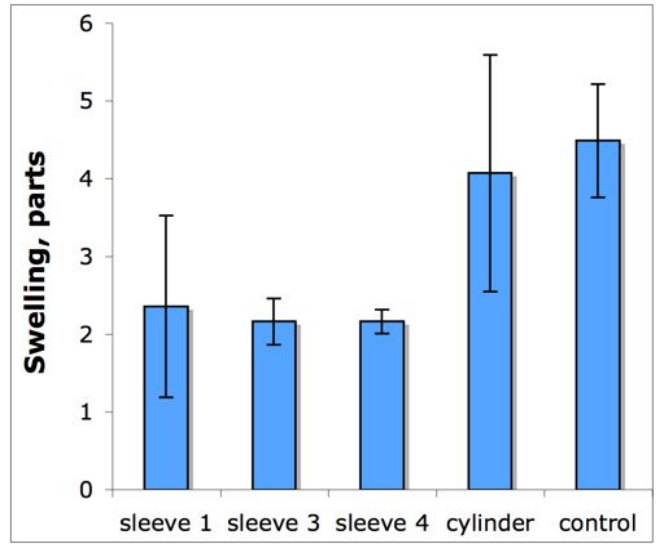

Fig.14. Swelling of uncured SE70 prepreg in acetone: sleeve 1-4 and cylinder samples are from the flight cassette, the control is from the ground cassette.

The swelling degree results correlate with gel-fraction content results (Fig.14). The epoxy matrix in the control samples swells to a higher degree, than the matrix of the flight samples on the sleeve. The swelling of the flight sample on the cylinder is the same as for the control sample. The higher content of gel-fraction and lower swelling degree in the flight samples shows, that the crosslinking is more dense in the flight samples, than in the ground control.

FTIR ATR spectra from the samples were recorded using a Digilab FTS7000 FTIR spectrometer fitted with an ATR accessory (Harrick, USA) with trapezium germanium crystal and incidence angle of 45°. To obtain sufficient signal/noise ratio and resolution of spectral bands, we used 500 scans and a resolution of 4 cm$^{-1}$. The absorbance of the 915 cm$^{-1}$ line of the epoxy group vibration was used for the analysis of the concentration of epoxy groups in the samples. In order to compare between spectra the absorbance was normalised to the absorbance of the 1509 cm$^{-1}$ line corresponding to aromatic ring vibrations, which is not affected by the curing. The FTIR ATR spectra of the samples were recorded on the 24 of April, 6 days after landing.

The absorbance of epoxy group vibrations shows, that the epoxy groups concentration in the uncured flight cassette samples and in the uncured ground control samples is similar (Fig.15). The absorbance of epoxy groups in both kinds of the samples is high and close to the absorbance of 0.22 for the initial uncured prepreg. It shows, that the curing reaction of epoxy groups with active groups of hardener in epoxy matrix occurred slowly in the flight and ground control samples due to the low temperatures experienced by the flight and ground control cassettes.

The FTIR ATR spectra of gel-fraction of the samples after washing in acetone show the residual epoxy groups, which are attributed to the crosslinked macromolecules of the epoxy resin (Fig.16). The low absorbance of residual epoxy groups in the gel-fraction of the ground control sample (0.038) is similar to the absorbance of residual epoxy groups in the sample cured before the flight (0.041). It shows that the gel-fraction in the ground control samples is formed due to the reaction of epoxy groups with hardener as in the usual curing of the epoxy resin composite.

The absorbance of residual epoxy groups in the gel-fraction of the flight samples is higher than in the gel-fraction of the ground control samples. This means that the high degree of crosslinking of the macromolecules in the flight samples is not due to the curing reaction of epoxy groups. There are clearly additional reactions, which can cause a crosslinking in the flight samples.

An important parameter of the uncured prepreg after exposure in the stratosphere is the remaining chemical activity for curing. The flight and control samples were cured in the thermobox at $80^0$C over a period of 3 days. The completeness of the curing reaction was tested by FTIR ATR spectroscopy (Fig.15). The FTIR ATR spectra of the cured samples show the absorbance of residual epoxy groups in the composite after the curing reaction. The absorbance of residual epoxy groups in the cured ground control samples is similar to the absorbance of epoxy groups in flight sample 5 which was cured before the flight. The absorbance of the residual epoxy groups in the cured flight samples is higher, than in the ground control samples. This means that the epoxy groups in the flight samples were not completely reacted during the curing. At the same time, the flight samples are solid, not sticky and appear similar to the completely cured control samples.

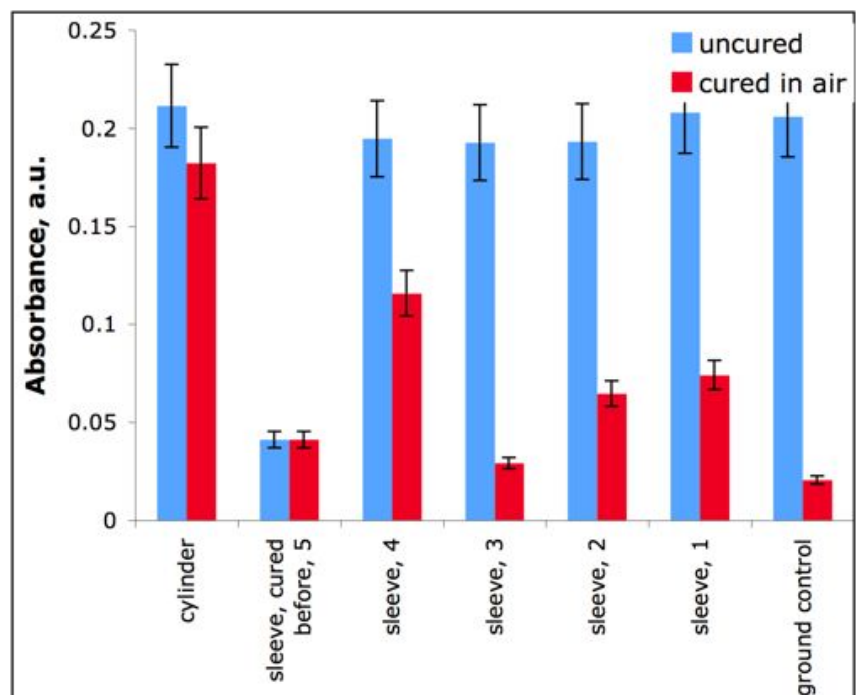

Fig.15. Normalized absorbance at 915 cm$^{-1}$ line in the FTIR ATR spectra of uncured SE70 prepreg and SE70 composite cured in air at $80^0$ C over 3 days: 1-4 sleeve samples and sample on the cylinder in the flight cassette; sleeve 5 is the flight cassette sample cured before the flight; control is the sample in the ground cassette.

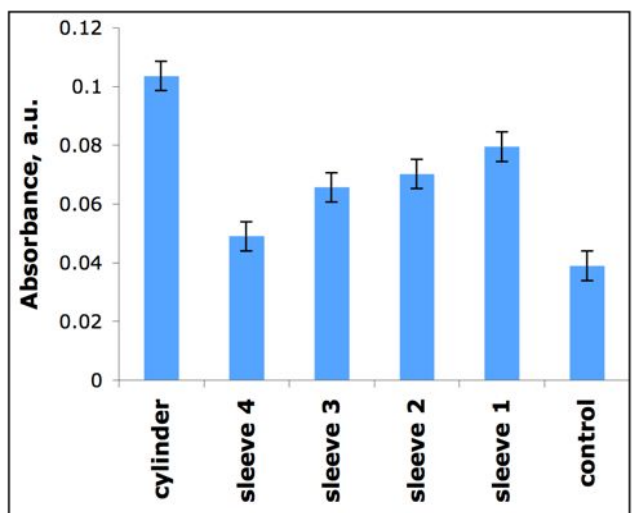

Fig.16. Absorbance at 915 cm$^{-1}$ from FTIR ATR spectra of the gel-fraction in the SE70 prepregs: 1-4 sleeve samples and the sample on cylinder are from the flight cassette, the control is the ground cassette.

**Discussion**

The flight samples remained uncured in stratosphere and were successfully cured after landing. The conditions in the stratosphere did not interfere with the curing capability of the samples. Therefore, the uncured samples can be delivered and stored in the stratosphere before curing.

The temperature records during the flight show, that the uncured samples were exposed to temperatures in the range from -76$^0$C to +32$^0$C. The prepreg used has a slow curing reaction rate at such low temperatures, as noted in the Gurit Information Sheet for the SE70 prepreg. The small decrease of the epoxy group concentration in the samples in the ground cassette and in the flight cassette from the original uncured sample shows that the curing reaction proceeded minimally in accordance with the low curing rate. The temperature of the ground cassette was higher than the temperature of the flight cassette, but the temperature difference was too small to observe a difference in epoxy group conversion between the ground control and flight cassette samples. However, the analysis of gel-fraction and swelling degree shows that the flight cassette samples have a significantly higher degree of crosslinking. The gel-fraction of the control samples is formed as the result of epoxy group conversation and this residual concentration is equal in the completely cured samples and the incompletely cured samples. In the flight samples, the concentration of residual epoxy groups is higher, indicating that the gel-fraction is formed due epoxy group reaction as well as additional reactions which leave the epoxy groups unchanged.

Therefore, the gel-fraction in the flight samples is due to two reactions: the reaction of epoxy groups with the active groups of the hardener, as in the usual curing process, and additional chemical reactions, where the epoxy groups do not take part. These additional reactions are initiated by radiation damage in epoxy matrix due to high energy particles (electrons and ions), X-rays, γ-rays, ultraviolet and short wavelength ultraviolet sunlight as per the specific conditions in the stratosphere. Space radiation can cause damage of the polymer macromolecules producing free radicals, which facilitate crosslinking between macromolecules without conversion of the epoxy groups of the macromolecules. The intensity of these space radiations at a height of

40 km in the stratosphere above the ozone layer is higher than on the ground, therefore, the flight samples were irradiated with a higher intensity than the ground control samples.

In summary, our results show that the stratospheric conditions give additional pathways for crosslinking of the polymer matrix. Similar effects of radiation damage on curing reactions were observed in our previous laboratory investigations in which prepregs were exposed to plasma and ion beam [9, 10, 17]. Similarly additional crosslinking reactions in uncured composites can be expected in a low Earth orbit free space environment.

**Conclusions**

This is the first time that uncured epoxy matrix composites were exposed in the stratosphere at a height of 40 km over 3 days. Analysis of the gel-fractions and epoxy group concentrations showed that curing reactions in the flight samples and in the ground control samples are different. Additional crosslinking reactions caused by space radiations are observed in the flight samples. The uncured samples remain curable after an exposure in stratosphere.

**Acknowledgments**

Authors thanks Mr. David Gregory (NASA Balloon Program) for excellent organization of the scientific stratospheric balloon program in Australia; Mr. William Stepp, Mr. Scott Hadley and their colleagues (Columbia Scientific Balloon Facility, CSBF) for outstanding launch, smooth flight of the balloon and gentle landing of the payload; Prof. Allen Zych and his colleagues for useful collaboration and friendly help with payload; Prof. Ravi Sood (ADFA) for support of the mission in Alice Springs; Dr. Miriam Baltuck (CSIRO Astronomy and Space Science), Mr. Bruce Banks, Dr. Viet Nguyen and Mrs. Kim de Groh (NASA Glenn Research Center) for collaboration and support of the mission; Prof. Anna Green and Mr. Paul Harbon (University of Sydney) for financial and organizational support of the mission; Mr. Neil Loveland and Mr. Paul Kulmar (Gurit, Australia) for the prepreg samples and useful discussions; Mr. Robert Davis (University of Sydney) for the preparation of the payload equipment; Mr. Paul Wilson (St. Phillip College, Alice Springs) for help with preparation of epoxy resin mixtures.

**References**

1. Kondyurin A.V., Building the shells of large space stations by the polymerisation of epoxy composites in open space, Int. Polymer Sci. and Technol., v.25, N4, 1998, p.T/78.
2. Briskman V., A.Kondyurin, K.Kostarev, V.Leontyev, M.Levkovich, A.Mashinsky, G.Nechitailo, T.Yudina, Polymerization in microgravity as a new process in space technology, Paper № IAA-97-IAA.12.1.07, 48th International Astronautical Congress, October 6-10, 1997, Turin Italy.
3. Kondyurin A., G.Mesyats, Yu.Klyachkin, Creation of High-Size Space Station by Polymerisation of Composite Materials in Free Space, J. of the Japan Soc. of Microgravity Appl., v.15, Suppl.II, 1998, p.61-65.

4. Kondyurin A., Kostarev K., Bagara M.V., Polymerization processes of epoxy plastic in free space conditions, Paper IAF-99-I.5.04, 50th International Astronautical Congress 4-8- Oct., 1999, Amsterdam, The Netherlands.
5. Kondyurin A., High-size space laboratory for biological orbit experiments, Advanced space research, v.28, N4, 2001, pp.665-671
6. Kondyurin A., Kostarev K., Bagara M., Polymerization processes of epoxy plastic in simulated free space conditions, Acta Astronautica, vol.48, N2-3, 2001, pp.109-113
7. Briskman V.A., Yudina T.M., Kostarev K.G., Kondyurin A.V., Leontyev V.B., Levkovich M.G., Mashinsky A.L., Nechitailo G.S., Polymerization in microgravity as a new process in space technology, Acta Astronautica, vol.48, N2-3, 2001, pp.169-180
8. Kondyurin A., B. Lauke, I. Kondyurina and E. Orba, Creation of biological module for self-regulating ecological system by the way of polymerization of composite materials in free space, Advances in Space Research, 2004, v. 34/7, p. 1585-1591
9. Kondyurin A., B.Lauke, E.Richter: Polymerization Process of Epoxy Matrix Composites under Simulated Free Space Conditions, High Performance Polymers. 16, 2004, p. 163 – 175
10. Kondyurin A., B.Lauke: Curing of liquid epoxy resin in plasma discharge, European Polymer Journal. 40/8, 2004, p. 1915 – 1923
11. Kondyurina I., A. Kondyurin, B. Lauke, L. Figiel, R. Vogel, U. Reuter, Polymerisation of composite materials in space environment for development of a Moon base, Advances in space research, 37, 2006, p.109-115
12. Kondyurin A., B. Lauke, R. Vogel, Photopolymerisation of composite material in simulated free space environment at low Earth orbital flight, European Polymer Journal 42 (2006) 2703–2714;
13. A. Kondyurin, B. Lauke, R. Vogel, G. Nechitailo, Kinetics of photocuring of matrix of composite material under simulated conditions of free space, Plasticheskie massi, 2007, v.11, pp.50-55
14. A.V.Kondyurin, L.A.Komar, A.L.Svistkov, Modelling of curing of composite materials for the inflatable structure of a lunar space base, Mechanics of composite materials and constructions, 15 (4), 512-526, 2009.
15. A.V.Kondyurin, G.S.Nechitailo, Composite material for Inflatable Structures Photocured under Space Flight Conditions, Cosmonautics and rockets, 3 (56), 182-190, 2009
16. A. Kondyurin, M. Bilek, Ion Beam Treatment of Polymers. Application aspects from medicine to space, Elsevier, Oxford, 2008;
17. A. Kondyurin, M. Bilek, Etching and structure transformations in uncured epoxy resin under rf-plasma and plasma immersion ion implantation, Nuclear Instruments and Methods in Physics Research, B 268 (2010) 1568–1580.